\documentclass[reprint,prl,twocolumn,superscriptaddress,showpacs,showkeys,floatfix,preprintnumbers]{revtex4-1}

\usepackage{braket}
\usepackage{xargs}
\usepackage{mathtools}
\usepackage[english]{babel}
\usepackage[utf8]{inputenc}
\usepackage{amsmath,amssymb,braket,slashed,mathrsfs}
\usepackage{pifont}
\usepackage{MnSymbol}
\usepackage{graphicx}
\usepackage{tikz}
\usetikzlibrary{shapes,arrows,positioning}
\tikzset{decision/.style={diamond, draw, fill=blue!20, text width=4.5em, text badly centered, inner sep=0pt}}
\tikzset{block/.style={rectangle, draw, fill=blue!20, text width=10em, text centered, rounded corners, minimum width=3.5cm}}
\tikzset{block1/.style={rectangle, draw, fill=blue!20, text width=18.5em, text centered, rounded corners, minimum width=3.5cm}}
\tikzset{line/.style={draw, -latex, thick}}
\usepackage{color,hyperref}
\usepackage{multirow,array}
\usepackage{physics}

\usepackage{cleveref}
\newcommand{\be}{\begin{equation}}
	\newcommand{\ee}{\end{equation}}
\newcommand{\ba}{\begin{eqnarray}}
\newcommand{\ea}{\end{eqnarray}}

\newcommand{\nn}{\nonumber}

\newcommand{\lat}{\mathrm{lat}}

\renewcommand{\vec}[1]{\mathbf{#1}}
\newcommand{\ud}{\mathrm{d}}

\newcommand{\innovation}{Collaborative Innovation Center of Quantum Matter, Beijing 100871, China}
\newcommand{\chep}{Center for High Energy Physics, Peking University, Beijing 100871, China}
\newcommand{\pkuphy}{School of Physics, Peking University, Beijing 100871,
	China}
\newcommand{\KeyLab}{State Key Laboratory of Nuclear Physics and Technology,
	Peking University, Beijing 100871, China}
\newcommand{\Uconn}{Department of Physics, University of Connecticut, Storrs, CT 06269, USA}
\newcommand{\RBRC}{RIKEN-BNL Research Center, Brookhaven National Laboratory, Building 510, Upton, NY 11973}

\begin{document}
	\title{Lattice QCD calculation of the two-photon exchange contribution to the muonic-hydrogen Lamb shift}
	
	\author{Yang~Fu}\affiliation{\pkuphy}
	\author{Xu~Feng}\email{xu.feng@pku.edu.cn}\affiliation{\pkuphy}\affiliation{\innovation}\affiliation{\chep}\affiliation{\KeyLab}
	\author{Lu-Chang Jin}\email{ljin.luchang@gmail.com}\affiliation{\Uconn}\affiliation{\RBRC}
	\author{Chen-Fei Lu}\affiliation{\pkuphy}
	%
	
	\date{\today}
	
	\begin{abstract}
		We develop a method for lattice QCD calculation of the two-photon exchange contribution to the muonic-hydrogen Lamb shift. To demonstrate its feasibility, we present the first lattice calculation with a gauge ensemble at $m_\pi = 142$ MeV. By adopting the infinite-volume reconstruction method along with an optimized subtraction scheme, we obtain $\Delta E_{\text{TPE}} = -28.9(4.9)~\mu\text{eV} + 93.72~\mu\text{eV}/\text{fm}^2 \cdot\langle r_p^2 \rangle$, or $\Delta E_{\text{TPE}} = 37.4(4.9)~\mu$\text{eV}, which is consistent with the
previous theoretical results in a range of 20-50 $\mu$eV.
	\end{abstract}
	
	\maketitle
	
	\section{Introduction}
\begin{figure}[htbp]
		\begin{center}
			\includegraphics[width=45mm]{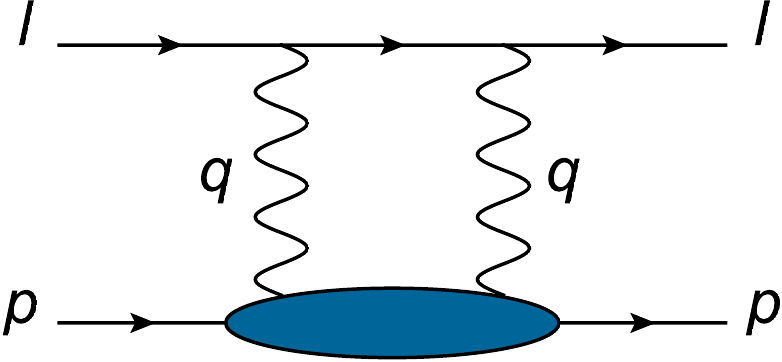}
			\caption{The diagram for two-photon exchange correction.}
			\label{fig:TPE}
		\end{center}
	\end{figure}

Historically, the discovery of hydrogen's Lamb shift laid the foundation for the modern quantum electrodynamics.
In recent years the precise measurements of muonic-hydrogen ($\mu$H) Lamb shift~\cite{Pohl:2010zza,Antognini:2013txn} yielded
the most precise determination of the proton charge radius,
but raised a 7~$\sigma$ discrepancy from the CODATA-2010 value~\cite{Mohr:2012tt}, known as the proton radius puzzle.
In 2019, two experiments reported results which agree with the $\mu$H measurements~\cite{Bezginov:2019mdi,Xiong:2019umf} and represented a decisive step towards solving the puzzle for a decade.
On the theoretical side, the puzzle has triggered many efforts to improve 
the theoretical understanding of both spectroscopy and scattering.
Among them, the two-photon exchange (TPE) contribution, see Fig.~\ref{fig:TPE}, is of special interest. 
It introduces the largest theoretical uncertainty to both the Lamb shift and hyperfine splitting in $\mu$H~\cite{Antognini:2013rsa}.
In addition, it plays an important role in extracting the charge radius from scattering experiments at high precision~\cite{Gorchtein:2014hla} and resolving the proton electric to magnetic form factor ratio puzzle~\cite{Arrington:2011dn}
induced by using the Rosenbluth separation~\cite{Rosenbluth:1950yq} and the polarization transfer methods~\cite{JeffersonLabHallA:1999epl}.
	
\begin{figure}[htbp]
                \begin{center}
                        \includegraphics[width=85mm]{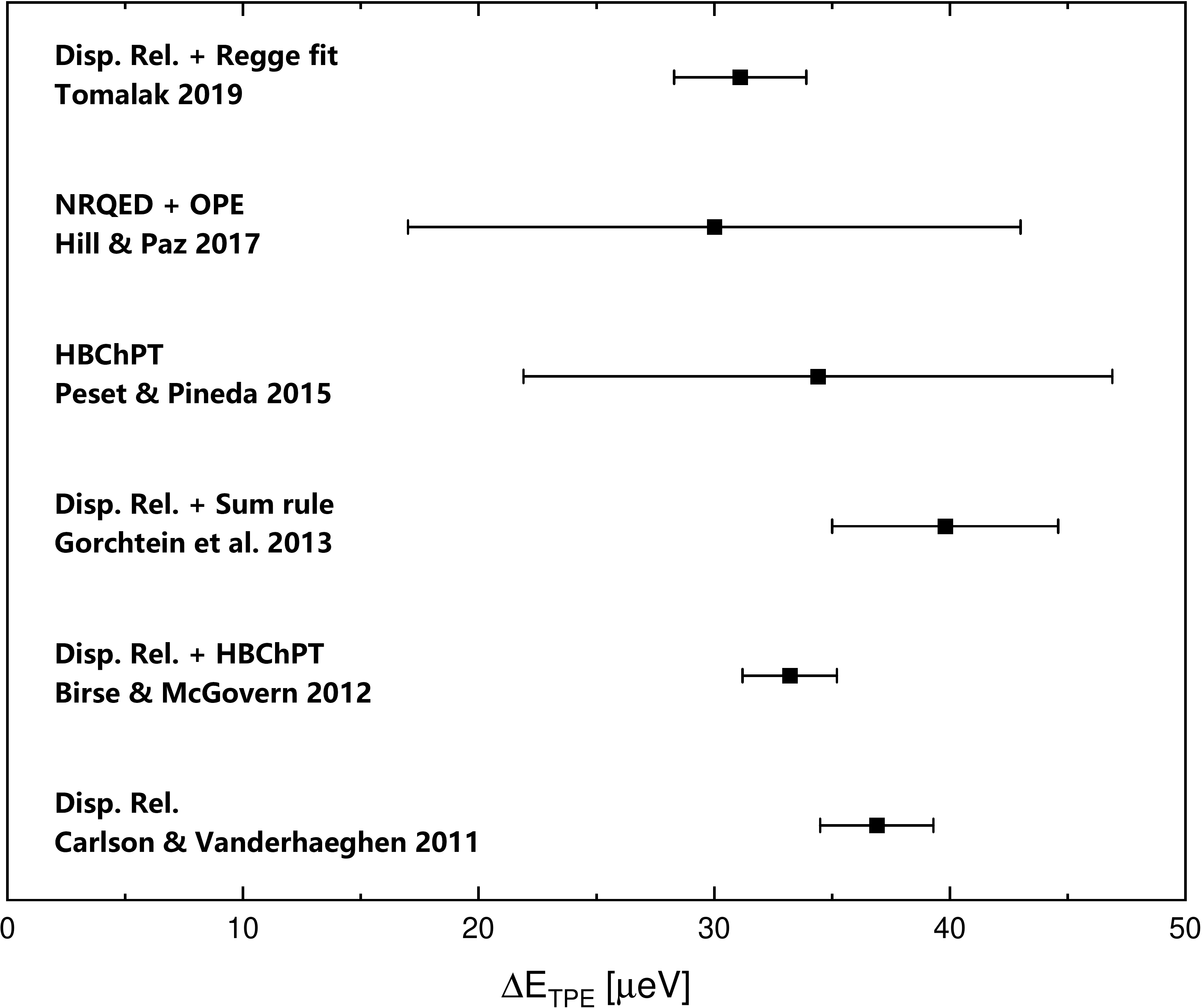}
                        \caption{Theoretical results for the TPE contribution. From top to bottom, the results are refered to Refs.~\cite{Tomalak:2018uhr},\cite{Hill:2016bjv},\cite{Peset:2015zga},\cite{Gorchtein:2013yga},\cite{Birse:2012eb} and \cite{Carlson:2011zd}, respectively.}
                        \label{fig:TPE_review}
                \end{center}
        \end{figure}

	Several approaches have been proposed to calculate the TPE correction to the $\mu$H Lamb shift, 
including dispersion relations (DR) \cite{Pachucki:1999zza,Martynenko:2005rc,Carlson:2011zd,Gorchtein:2013yga} , baryon $\chi$PT (B$\chi$PT) \cite{Alarcon:2013cba,Alarcon:2020wjg}, heavy baryon $\chi$PT (HB$\chi$PT) \cite{Nevado:2007dd,Birse:2012eb,Peset:2015zga}, non-relativistic QED (NRQED) \cite{Hill:2011wy} and operator product expansion (OPE) \cite{Hill:2016bjv}. For these methods, the TPE correction is usually divided into Born and non-Born pieces, where the Born part is well-constrained by the experimental data, but the non-Born part contains a 
subtraction function, which is poorly constrained and relies on model, thus leading to a large systematic uncertainty. 
It is proposed recently that the subtraction function can be further constrained by the dilepton electroproduction~\cite{Pauk:2020gjv}. To date, the theoretical results of the TPE correction $\Delta E_\text{TPE}$ are summarized in Fig.~\ref{fig:TPE_review}.
These results are rather consistent but still vary in a range of 20-50 $\mu$eV.

The total $2S-2P$ Lamb shift is given by~\cite{Antognini:2013rsa} 
	\be \label{eq:muH_LS_theory}
	\Delta E_\text{LS}^{\text{theory}} = 206033.6(1.5) - 5227.5(1.0) \langle r_p^2 \rangle + \Delta E_\text{TPE}.
	\ee
with $\langle r_p^2 \rangle$ the square of the charge radius. 
In Eq.~(\ref{eq:muH_LS_theory}) and through out the paper we assume radii to be in fm, resulting energies in $\mu$eV.
Using $\Delta E_\text{TPE} = 33.2(2.0)$ $\mu$eV from Ref.~\cite{Birse:2012eb}
and the experimental value $\Delta E_\text{LS}^{\text{exp}} = 202370.6(2.3)$ $\mu$eV, the charge radius $\sqrt{\langle r_p^2\rangle}=0.84087(39)$ fm
is obtained, which causes the radius puzzle~\cite{Antognini:2013txn}. 
To resolve the puzzle, $\Delta E_\text{TPE}$ is required to be $\sim$300 $\mu$eV, 10 times larger than the current theoretical results.
Although $\Delta E_\text{TPE}$ is unlikely responsible for such large discrepancy, it contributes the largest theoretical uncertainty in extracting the charge radius
from $\mu$H Lamb shift.
Any further improvement on our understanding of the proton size would inevitably require an improved determination of 
$\Delta E_\text{TPE}$,
particularly from lattice QCD to avoid the uncertainties induced by model assumptions.

	Recently several lattice QCD approaches have been proposed, including using the Feynman-Hellmann theorem to calculate the Compton amplitude~\cite{Can:2020sxc} and using a different subtraction point to calculate the subtraction function~\cite{Hagelstein:2020awq}.
In this work, we develop a method to directly calculate the whole TPE contribution to $\mu$H Lamb shift and perform a realistic lattice calculation with a gauge ensemble at the pion mass $m_\pi=142$ MeV.
	
	\section{Two-photon exchange contribution}

We start with the spin-averaged forward doubly-virtual Compton scattering tensor in Euclidean space
	\begin{align} \label{eq:Compton_tensor}
		\mathcal{T}_{\mu\nu}(P,Q) &= \frac{1}{8\pi M} \int \ud^4 x e^{iQ \cdot x} \langle p \vert \operatorname{T}[j_\mu(x) j_\nu(0)] \vert p \rangle \nn \\[1ex]
		& = \left( -\delta_{\mu \nu} + \frac{Q_\mu Q_\nu}{Q^2} \right)  \mathcal{T}_1(\nu, Q^2) \nn \\
		&- \left( P_{\mu} - \frac{P \cdot Q}{Q^2} Q_{\mu}\right)\left( P_{\nu} - \frac{P \cdot Q}{Q^2} Q_{\nu}\right) \frac{\mathcal{T}_2(\nu, Q^2)}{M^2} ,
	\end{align}
	where $\nu = P\cdot Q/M$ with $P= (iM,\vec{0})$ and $Q=(Q_0,\vec{Q})$ the Euclidean proton and photon four-momenta. $M$ is the proton mass and $j_{\mu,\nu}$ are the electromagnetic quark currents. 	

The relative energy shift to the $nS$-state is given by \cite{Pachucki:1999zza}
	\begin{align} \label{eq:TPE}
		E &=  \frac{8 m \alpha^2}{ \pi } \abs{\phi_n(0)}^2 \int \ud^4Q 
		\nn \\[1ex]
		&\times \frac{ (Q^2+2Q_0^2) \mathcal{T}_1(iQ_0,Q^2) - (Q^2 -Q_0^2) \mathcal{T}_2(iQ_0,Q^2) }
		{ Q^4 (Q^4 + 4m^2 Q_0^2) },
	\end{align}
	where $m$ is the lepton mass and $\abs{\phi_n(0)}^2 = {m_r^3 \alpha^3}/{(\pi n^3)}$ is the square of the $nS$-state wave function
at the origin with $m_r = mM/(M+m)$ the reduced mass and $\alpha$ the fine structure constant. Note that the $nP$-state wave function vanishes at the origin, hence it does not
 receive any corrections from TPE at this order. 

The TPE correction to the $\mu$H Lamb shift has no infrared (IR) divergence because the binding energy serves as an IR regulator. 
By treating the proton as a point-like particle with the corrections from charge radius~\cite{Carlson:2011zd}
one can calculate such contribution based on bound-state QED~\cite{Pachucki:1996zza,Pachucki:1999zza,Borie:2004fv}.
However, more precise comparison between 
theory and experiment is limited due to the lack of the information on the proton structure. The energy shift $E$ defined 
in Eq.~(\ref{eq:TPE}) contains all the required structure information, but is unfortunately IR divergent 
as the lepton in the Compton scattering is no more bounded. 
Here, the idea is to obtain the structure-dependent TPE correction from Eq.~(\ref{eq:TPE}) 
by subtracting the contributions from a point-like proton and 
the third Zemach moment~\cite{Carlson:2011zd}.
For the former, the contribution can be described by
the proton-photon vertex function $\Gamma_\mu = \gamma_\mu$ with
        \be \label{eq:sub1}
        \mathcal{T}_{1}^{pt} = \frac{M}{\pi} \frac{\nu^2}{Q^4 - 4M^2 \nu^2},\quad \mathcal{T}_{2}^{pt} =  \frac{M}{\pi} \frac{Q^2}{Q^4 - 4M^2 \nu^2}.
        \ee
For the latter, it is given by
        \be \label{eq:sub2}
        E^{Z} = - \alpha^2 \abs{\phi_n(0)}^2 \int \frac{\ud Q^2}{Q^2} \frac{16mM}{(M+m)Q} G_E^\prime(0) ,
        \ee
where $G_E(Q^2)$ is the proton electric form factor and its derivative is related to proton charge radius via $\langle r_p^2 \rangle = -6\,G_E^\prime(0)$. 
After the subtraction of $E^{pt}$ and $E^Z$, one obtains the IR-finite TPE contribution $\Delta E=E-E^{pt}-E^Z$.
The quantity $\Delta E_{\text{TPE}}$ shown in Eq.~(\ref{eq:muH_LS_theory}) is defined as the difference between energy shifts to the $nP$- and $nS$-states and thus we have $\Delta E_{\text{TPE}}=-\Delta E$. 

For a realistic lattice QCD calculation, Eq.~(\ref{eq:TPE}) cannot be used directly as it is IR divergent. Thus the first difficulty we
need to conquer is to write $\Delta E_\text{TPE}$ in terms of the hadronic functions calculable via lattice simulations and maintain the IR cancellation automatically.
	
	\section{Lattice QCD methodology}
	
	On the lattice, we prefer to rewrite Eq.(\ref{eq:TPE}) in terms of $T_1=\mathcal{T}_{00}$ and $T_2=\sum_i \mathcal{T}_{ii}$ as
	\be \label{eq:TPE_lat}
	E = -16 m \alpha^2 \abs{\phi_n(0)}^2 \int_{\varepsilon} \frac{\ud Q^2}{Q^4}\int_{-\frac{\pi}{2}}^{\frac{\pi}{2}} \ud\theta\,(\alpha_1 T_1+\alpha_2 T_2),
	\ee
with 
\be
\alpha_1(Q)=\frac{1-\sin^4\theta}{1 + \sin^2\theta/\tau_\ell},\quad \alpha_2(Q)=\frac{\sin^2\theta(1-\sin^2\theta)}{1 +\sin^2\theta/\tau_\ell}
\ee
and $\tau_\ell=\frac{Q^2}{4m^2}$.
Here the angle $\theta$ is defined as $Q_0=Q\sin\theta$ and $|\vec{Q}|=Q\cos\theta$.
The notation $\int_{\varepsilon}$ indicates that the integral is performed in the region of $Q^2\ge \varepsilon^2$ with an IR regulator $\varepsilon$.

	Combining Eqs.~(\ref{eq:Compton_tensor}) and (\ref{eq:TPE_lat}), we obtain
	\be \label{eq:master_formula_naive}
	E = \frac{2 m \alpha^{2}}{\pi M} \abs{\phi_n(0)}^2 \sum_{i=1,2}\int \ud^{4} x ~ \bar{\omega}_{i}(\vec{x}, t) H_{i}(\vec{x}, t),
	\ee
	where the hadronic functions are defined as
	\begin{align}
		H_{1}(\vec{x}, t) &=\langle p|\operatorname{T}[j_{0}(\vec{x}, t) j_{0}(0)]| p\rangle , \nn \\[1ex]
		H_{2}(\vec{x}, t) &=\langle p|\operatorname{T}[\vec{j}(\vec{x}, t) \cdot \vec{j}(0)]| p\rangle.
	\end{align}
	The weight functions $\bar{\omega}_{i}(\vec{x}, t)$ are given by
	\begin{equation}
\label{eq:weight_function}
		\bar{\omega}_{i}(\vec{x}, t)=-\int_{\varepsilon} \frac{\ud Q^{2}}{Q^{4}} \int_{-\frac{\pi}{2}}^{\frac{\pi}{2}} \ud \theta 
\,\alpha_i(Q)f(Q;x)
	\end{equation}
with
\begin{equation}
f(Q;x)=\cos(Q_0t)j_0(|\vec{Q}||\vec{x}|).
\end{equation}
Here an average over the spatial directions is taken and $j_n(x)$ are the spherical Bessel functions. 

Using the infinite-volume reconstruction method \cite{Feng:2018qpx},
we split the time integral in Eq.~(\ref{eq:master_formula_naive}) into the regions $|t|<t_s$ and $|t|\ge t_s$ and have
\be
E=E^{<t_s}+E^{\ge t_s}.
\ee 
At sufficiently large $t_s$, ground-state dominance allows us to relate $H_i(\vec{x},t)$ at $|t|\ge t_s$
to $H_i(\vec{x},t_s)$. Thus, $E^{\ge t_s}$ can be written as
\be
E^{\ge t_s}=\frac{2 m \alpha^{2}}{\pi M} \abs{\phi_n(0)}^2\sum_{i=1,2} \int \ud^{3}\vec{x} ~ \bar{L}_{i}(\vec{x}, t_s) H_{i}(\vec{x}, t_s),
\ee
where the weight function $\bar{L}_{i}$ is defined as
\be
\label{eq:weight_function_LD}
\bar{L}_i(\vec{x},t_s)=
-\int_{\varepsilon}\frac{\ud Q^2}{Q^4}\int_{-\frac{\pi}{2}}^{\frac{\pi}{2}} \ud\theta\,\alpha_i(Q)g(Q;\vec{x},t_s)
\ee
with
\be
g(Q;\vec{x},t_s)=\frac{1}{M}\frac{A(Q,t_s)}{\tau_p+\sin^2\theta}j_0(|\vec{Q}||\vec{x}|),\quad \tau_p=\frac{Q^2}{4M^2},
\ee
and
\ba
&&A(Q,t_s)=A_c(Q)\cos(Q_0t_s)-A_s(Q)\sin(Q_0t_s),
\nn\\
&&A_c(Q)=\left(1/4+\tau_p\cos^2\theta\right)^{\frac{1}{2}}+1/2-\sin^2\theta,
\nn\\
&&A_s(Q)=\frac{\sin\theta}{\sqrt{\tau_p}}\left[\left(1/4+\tau_p\cos^2\theta\right)^{\frac{1}{2}}+1/2+\tau_p\right].
\ea

We originally hope that the IR divergent part is isolated by the large-$t$ contribution $E^{\ge t_s}$ and thus only
the weight function $\bar{L}_i$ is divergent when $\varepsilon\to0$. 
However, the situation is more complicated than expected as the small-$t$ contribution $E^{< t_s}$ 
is also IR divergent. 
(Although associated with $H_i(\vec{x},t)$ at small $t$, $\bar{\omega}_i$ receives 
significant long-distance contributions from the leptonic part
and thus is IR singular.)
To solve this difficulty, we split the weight functions into two parts
\be
\bar{\omega}_i=\hat{\omega}_i+\delta\omega_i,\quad \bar{L}_i=\hat{L}_i+\delta L_i,
\ee
where the divergent part is absorbed by $\delta\omega_i$ and $\delta L_i$ with
\ba
&&\delta\omega_i=-\int_{\varepsilon} \frac{\ud Q^{2}}{Q^{4}} \int_{-\frac{\pi}{2}}^{\frac{\pi}{2}} \ud \theta
\,\alpha_i(Q),
\nn\\
&&\delta L_i=-\int_{\varepsilon}\frac{\ud Q^2}{Q^4}\int_{-\frac{\pi}{2}}^{\frac{\pi}{2}} \ud\theta\,\alpha_i(Q)g_0(Q;\vec{x},t_s),
\ea 
and
\be
g_0(Q;\vec{x},t_s)=\frac{1}{M}\frac{\cos^2\theta}{\tau_p+\sin^2\theta}-2t_s.
\ee
One can confirm that $\hat{\omega}_i$ and $\hat{L}_i$ are now IR finite.
Accordingly, the energies $E^{<t_s}$ and $E^{\ge t_s}$ are written as
\be
E^{<t_s}=\hat{E}^{<t_s}+\delta E^{<t_s},\quad E^{\ge t_s}=\hat{E}^{\ge t_s}+\delta E^{\ge t_s}.
\ee

To determine $\delta E^{<t_s}$, we split $T_i(Q)$ into the small-$t$ and large-$t$ parts
\ba
T_i(Q)&=&\frac{1}{8\pi M}\int_{-t_s}^{t_s}\ud t\int\ud^3\vec{x}\,f(Q;x)H_i(\vec{x},t)
\nn\\
&+&\frac{1}{8\pi M^2}\frac{A(Q,t_s)}{\tau_p+\sin^2\theta}\tilde{H}_i(\vec{Q},t_s),
\ea
where $\tilde{H}_i(\vec{Q},t_s)$ is Fourier transformation of $H_i(\vec{x},t_s)$.
By assuming the ground-state dominance at $t_s$, 
$\tilde{H}_i(\vec{Q},t_s)$ is given by
\ba
\tilde{H}_i(\vec{Q},t_s)&=&\frac{M}{E_{\vec{Q}}}e^{-(E_{\vec{Q}}-M)t_s}
\\
&&\times
\begin{cases}
(E_{\vec{Q}}+M)G_E^2(Q_{\mathrm{on}}^2), & i=1,\\
-(E_{\vec{Q}}-M)\left[G_E^2+2G_M^2\right](Q_{\mathrm{on}}^2), & i=2,
\end{cases}
\nn
\ea
where $G_{E/M}$ are the proton electric and magnetic form factors with $Q_{\mathrm{on}}^2=2M(E_{\vec{Q}}-M)$ and $E_{\vec{Q}}=\sqrt{M^2+\vec{Q}^2}$.
On the other hand, $T_i(Q)$ can be written as a combination of Born and non-Born terms: $T_i(Q)=T_i^B(Q)+T_i^{NB}(Q)$,
where the Born term represents the elastic box and crossed box contributions and its analytical form is known~\cite{Carlson:2011zd,Birse:2012eb,Gasser:2015dwa}.
For the non-Born term, as it does not contain the pole structure, Ward identity requires it to vanish 
as $Q\to0$~\cite{Gasser:2015dwa}. By requiring $\lim_{Q\to0} T_i(Q)-T_i^B(Q)=0$, we obtain with large $t_s$
\be
\label{eq:low_mom_relation}
K_i=\frac{1}{2M}\int_{-t_s}^{t_s}\ud t\int \ud^3\vec{x}\,H_i(\vec{x},t)=
\begin{cases}
2\,t_s,&i=1,\\
\frac{3}{M},&i=2.
\end{cases}
\ee
This relation allows us to rewrite $\delta E^{<t_s}$ as
\be
\delta E^{<t_s}=-\frac{4 m \alpha^{2}}{\pi} \abs{\phi_n(0)}^2\sum_{i=1,2}\int_\varepsilon\frac{\ud Q^2}{Q^4}\int_{-\frac{\pi}{2}}^{\frac{\pi}{2}}\ud\theta\, \alpha_i(Q) K_i.
\ee

The last step is to perform the subtraction of 
$\Delta E=E-E^{pt}-E^Z$ as mentioned earlier. 
Here $E^{pt}$ can be calculated using the point-like proton contributions
\be
T_1^{pt}=\frac{M}{\pi}\frac{Q^2-Q_0^2}{Q^4+4M^2Q_0^2},\quad T_2^{pt}=\frac{M}{\pi}\frac{3Q_0^2}{Q^4+4M^2Q_0^2}.
\ee
The same IR regulator $\varepsilon$ shall be introduced to make $E^{pt}$ and $E^Z$ finite. 
One can relate the form factor $G_E(0)$ and the charge radius $\langle r_p^2\rangle$ to the hadronic function
as
\ba
&&G_E^2(0)=\int \ud^3\vec{x}\,L_0(\vec{x},t_s)H_1(\vec{x},t_s),
\nn\\
&&\langle r_p^2\rangle=\int \ud^3\vec{x}\,L_r(\vec{x},t_s)H_1(\vec{x},t_s),
\ea
with $t_s$ sufficiently large for ground-state dominance and
\be
L_0(\vec{x},t_s)=\frac{1}{2M},\quad L_r(\vec{x},t_s)=\frac{1}{4M}\left(\vec{x}^2-\frac{3+6Mt_s}{2M^2}\right).
\ee
These relations allow us to write $E^{pt}$ and $E^Z$ as an integral of $H_1(\vec{x},t_s)$.
Finally, we obtain
\ba
\label{eq:master_formula}
\Delta E &=& \frac{2 m \alpha^{2}}{\pi M} \abs{\phi_n(0)}^2\left\{ \sum_{i=1,2}\left[\int_{-t_s}^{t_s} \ud^{4} x ~ \omega_{i}(\vec{x}, t) H_{i}(\vec{x}, t)\right.\right.
\nn\\
&&\hspace{2.5cm}\left.+\int \ud^{3}\vec{x} ~ L_i(\vec{x}, t_s) H_{i}(\vec{x}, t_s)\right]
\nn\\
&&\left.-2M\int_\varepsilon\frac{\ud Q^2}{Q^4}\int\ud\theta\, \alpha_2(Q) \left(K_2-4\pi T_2^{pt}\right)\right\},
\ea
where $\omega_{i}(\vec{x}, t)=\hat{\omega}_i(\vec{x}, t)$, $L_2(\vec{x},t_s)=\hat{L}_2(\vec{x},t_s)$ and 
\ba
\label{eq:weight_function_LD_modified}
L_1(\vec{x},t_s)&=&
-\int_{\varepsilon}\frac{\ud Q^2}{Q^4} \bigg\{\int_{-\frac{\pi}{2}}^{\frac{\pi}{2}} \ud\theta\,\alpha_1(Q)[g-g_0](Q;\vec{x},t_s)
\nn\\
&&\hspace{1.5cm}+\frac{4\pi M^2Q}{3(M+m)}L_r(\vec{x},t_s)\bigg\}.
\ea
After the IR cancellation, the limit $\varepsilon\to0$ can be taken for Eqs.~(\ref{eq:master_formula}) and (\ref{eq:weight_function_LD_modified}) now.
The third line of Eq.~(\ref{eq:master_formula}) does not depend on $H_i(\vec{x},t)$ and thus can be calculated directly. It contributes
$-0.60$ $\mu$eV to $\Delta E$.

	\section{Optimized subtraction scheme}
The method described above 
provides a direct way to calculate the TPE contribution using hadronic functions $H_i(\vec{x}, t)$ as input, but it suffers from both the finite-volume effects and the signal-to-noise problem due to the fact that $L_1(\vec{x}, t_s)$ increases rapidly as the spatial distance $\abs{\vec{x}}$ increases. 
To solve this difficulty, we define a reduced weight function $L_1^{(r)}(\vec{x}, t_s)$ via
        \be \label{eq:L1_sub}
L_1^{(r)}(\vec{x}, t_s)=L_1(\vec{x}, t_s)-c_0 L_0(\vec{x}, t_s) - c_r L_r(\vec{x}, t_s).
        \ee
After the replacement of $L_1\to L_1^r$, the energy shift $\Delta E$ is now given by
\be
\Delta E = - 0.60 +\frac{2m\alpha^2}{\pi M}\abs{\phi_n(0)}^2 \left(c_0 + c_r \langle r_p^2 \rangle\right) + \Delta E_{\lat},
\ee
with $\Delta E_{\lat}$ computed using the first two lines of Eq.~(\ref{eq:master_formula}) but with $L_1$ replaced by $L_1^{(r)}$.

	\begin{figure}[htbp]
                \begin{center}
                        \includegraphics[width=85mm]{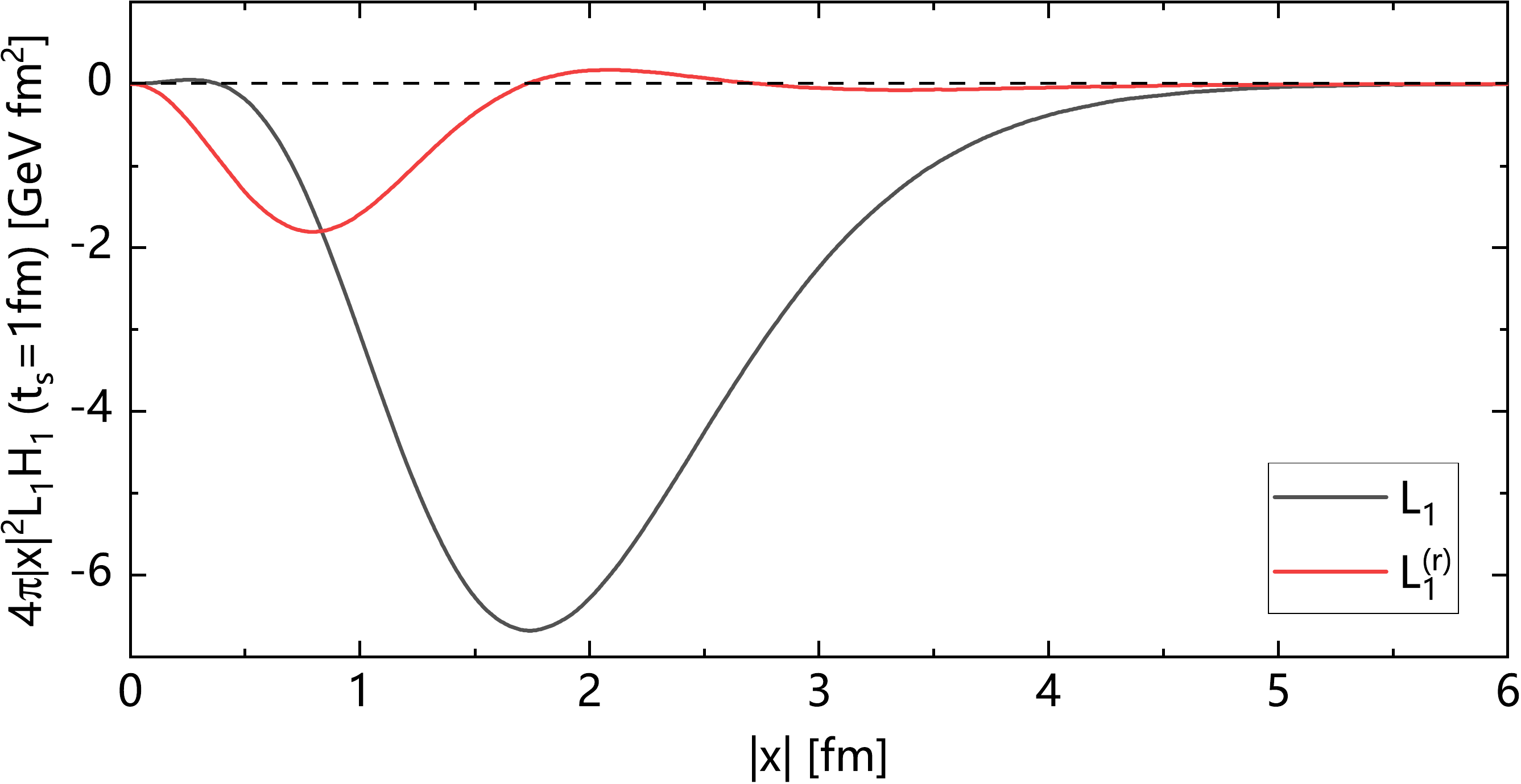}
                        \caption{The distributions of $4 \pi \abs{\vec{x}}^2 L_1(\vec{x}, t_s) H_1(\vec{x}, t_s)$ and 
$4 \pi \abs{\vec{x}}^2 L_1^{(r)}(\vec{x}, t_s) H_1(\vec{x}, t_s)$ at $t_s = 1$ fm, estimated using the dipole form factor.}
                        \label{fig:L1H1_model}
                \end{center}
        \end{figure}

Using the least squares method,
we determine the coefficients $c_0$ and $c_r$ by minimizing the following integral
\be
\label{eq:integral_minimization}
I(c_0,c_r)=\int_{R_{\text{min}}}^{R_{\text{max}}}\ud |\vec{x}|\,\left(4\pi|\vec{x}|^2\right) \abs{L_1^{(r)}(\vec{x}, t_s)}^2.
\ee
Here $c_0$ and $c_r$ depend on $t_s$, $R_{\text{min}}$ and $R_{\text{max}}$.
By examining the charge conservation from $\tilde{H}_1(\vec{0},t_s)$ and
the vanishing behavior of $\tilde{H}_2(\vec{0},t_s)$, we find the ground-state saturation at $t_s\gtrsim 0.8$ fm. 
In Eq.~(\ref{eq:integral_minimization}) we set $t_s=1$ fm.
By using a dipole functional form $G_E(Q^2) = 1/(1+ Q^2 \langle r_p^2 \rangle /12)^2$ with $\sqrt{\langle r_p^2 \rangle} = 0.85$ fm,
we mimic the distributions of $4\pi|\vec{x}|^2L_1(\vec{x}, t_s)H_1(\vec{x},t_s)$
and $4\pi|\vec{x}|^2L_1^{(r)}(\vec{x}, t_s)H_1(\vec{x},t_s)$ in Fig.~\ref{fig:L1H1_model}.
For the former, the main contribution comes from the range of 1-3 fm and the saturation occurs at $\sim5$ fm, requiring 
a large spatial volume with the size $L \simeq 10$ fm.
For the latter,
we set $R_\text{min}=1$ fm and $R_\text{max}=3$ fm and obtain
\ba
\label{eq:c0_cr}
&&\frac{2m\alpha^2}{\pi M}\abs{\phi_n(0)}^2\,c_0 = -0.17~\mu\text{eV},
\nn\\
&&\frac{2m\alpha^2}{\pi M}\abs{\phi_n(0)}^2\,c_r = -93.72~\mu\text{eV}/\text{fm}^2.
\ea
With the replacement $L_1\to L_1^{(r)}$, 
now the large-$|\vec{x}|$ contribution is significantly reduced and the saturation occurs at $\sim2.5$ fm.
With the values of $c_0$ and $c_r$ given in Eq.~(\ref{eq:c0_cr}), $\Delta E_{\text{TPE}}$ is given by
        \be \label{eq:TPE_split}
        \Delta E_{\text{TPE}} =-\Delta E= 0.77+ 93.72 \cdot\langle r_p^2 \rangle - \Delta E_{\lat}.
        \ee

	\section{Numerical results}
\begin{figure}[htbp]
		\begin{minipage}[h]{0.45\linewidth}
			\centerline{ \includegraphics[width=75pt]{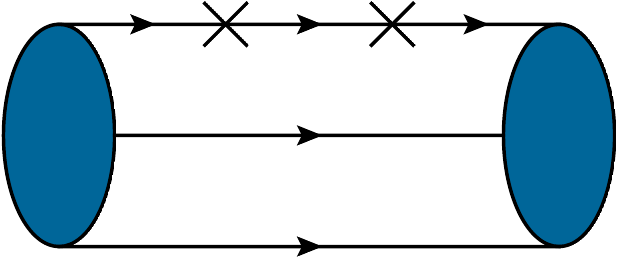} }
			Type I
		\end{minipage}
		\begin{minipage}[h]{0.45\linewidth}
			\centerline{ \includegraphics[width=75pt]{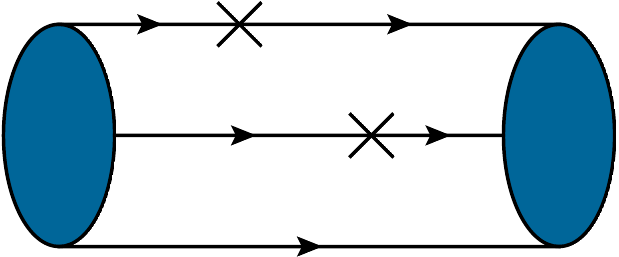} }
			Type II
		\end{minipage}
		\begin{minipage}[h]{0.32\linewidth}
			\centerline{ \includegraphics[width=75pt]{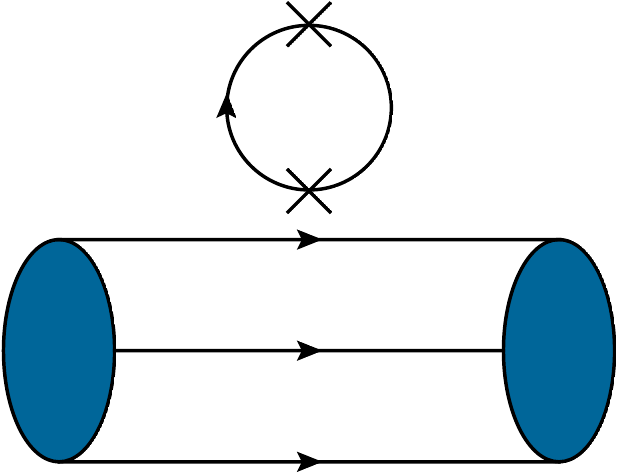} }
			Type III
		\end{minipage}
		\begin{minipage}[h]{0.32\linewidth}
			\centerline{ \includegraphics[width=75pt]{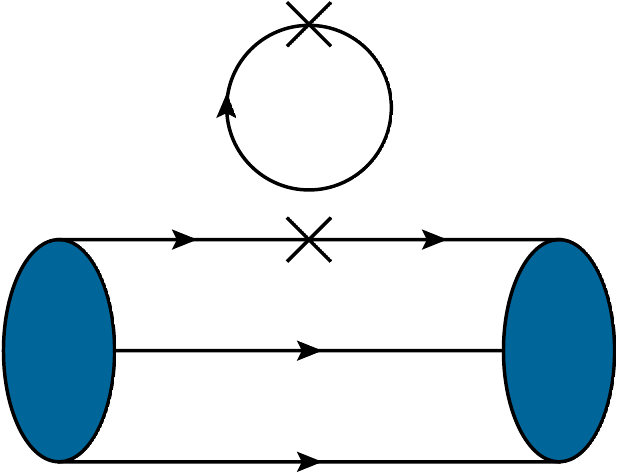} }
			Type IV
		\end{minipage}
		\begin{minipage}[h]{0.32\linewidth}
			\centerline{ \includegraphics[width=75pt]{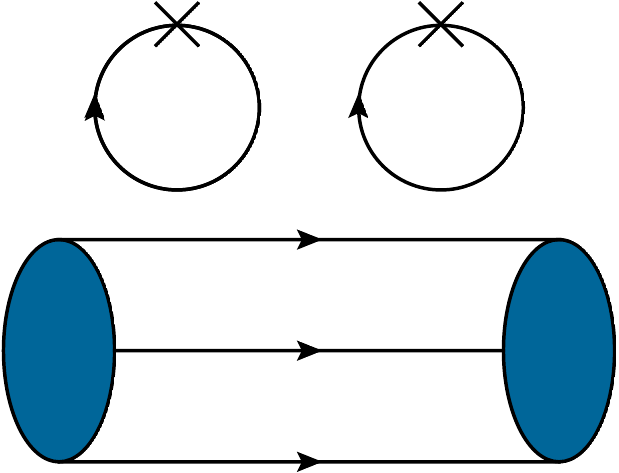} }
			Type V
		\end{minipage}
		\caption{Five types of quark field contractions for TPE diagrams. The blob denotes a proton state.}
		\label{fig:contractions}
	\end{figure}

 To demonstrate the feasibility of the methodology, we use a single gauge ensemble near the physical point, generated by the RBC-UKQCD Collaboration using $2+1$-flavor domain wall fermion~\cite{Blum:2014tka}. Ensemble parameters are listed in Table \ref{tab:ensemble_parameter}. We calculate the four-point correlation function 
$\sum_{\vec{x}_f, \vec{x}_i}\mathcal{P} \langle \psi_p(\vec{x}_f,t_f) j_\mu(x) j_\nu(y) \psi_p^\dagger(\vec{x}_i,t_i)\rangle$ using the random field sparsening technique \cite{Detmold:2019fbk,Li:2020hbj}, with the projection matrix $\mathcal{P} = (1+\gamma_0)/2$ and the time slices chosen as $t_i = \min\{t_x, t_y\} - \Delta t_i$ and $t_f = \max\{t_x, t_y\} + \Delta t_f$. 
The time separation $\Delta t_{i/f}$ should be sufficiently large for the proton ground-state saturation. In practice, we use 6 sets of $\{\Delta t_i/a,\Delta t_f/a\}=\{1,2\},\{2,1\},\{2,2\},\{2,3\},\{3,2\},\{3,3\}$ to examine the excited-state contamination for the initial/final state
and use $t_s/a=2,3,4,5$ to confirm the ground-state dominance for the intermediate state.
The total source-sink time separation ranges from 1.0 fm to 2.1 fm.
We use the local vector current $j_\mu$ with the renormalization factor 
quoted from Ref.~\cite{RBC:2014ntl} and further confirmed by our examination of the charge conservation.
 The quark field contractions for the TPE diagrams as shown in Fig.~\ref{fig:contractions}, with the first two the connected diagrams and the last three disconnected ones. 
We calculate both connected and disconnected diagrams with only 
Type IV and V neglected since they vanish in the flavor SU(3) limit. 
	
	\begin{table}[htbp]
		\normalsize
		\centering
		\begin{tabular}{ccccccc}
			\hline
			\hline
			Ensemble & $m_\pi$ [MeV] & $L/a$ & $T/a$ & $a$ [fm]&
			$N_{\text{conf}}$ \\
			\hline
			24D  & 142 & $24$ & $64$ & 0.1943(8) & 131 \\
			\hline
		\end{tabular}%
		\caption{Ensemble information. We list the pion mass $m_\pi$, the spatial and temporal extents, $L$ and $T$, the lattice spacing $a$, and the number of configurations used $N_{\text{conf}}$.}
		\label{tab:ensemble_parameter}%
	\end{table}
	
\begin{figure}[htbp]
		\begin{center}
			\includegraphics[width=85mm]{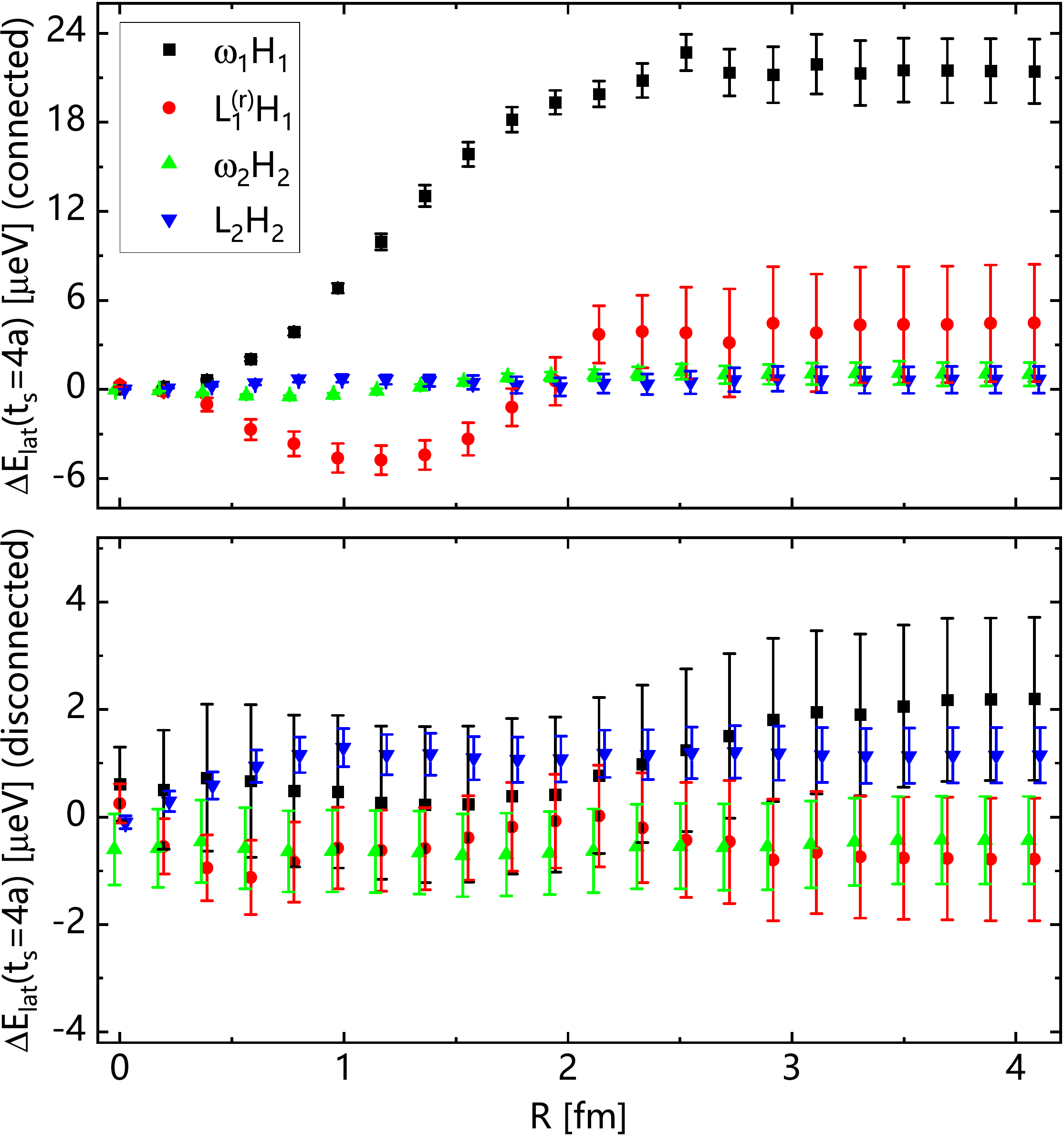}
			\caption{Results of $\Delta E_{\lat}$ as a function of the integral range $R$ at $\{\Delta t_i,\Delta t_f,t_s\}=\{2a,2a,4a\}$. The upper and lower panels show the results for the connected and disconnected contribution, respectively. Results from different terms have been slightly shifted for clarity.}
			\label{fig:TPE_ts4}
		\end{center}
	\end{figure}

	Using the lattice data at $\{\Delta t_i,\Delta t_f,t_s\}=\{2a,2a,4a\}$ as an example, in Fig.~\ref{fig:TPE_ts4} we show the results of $\Delta E_{\lat}$ as a function of the spatial integral range $R$.
	All four contributions to $\Delta E_{\lat}$ converge at large $R$ for both connected and disconnected diagrams, suggesting
that the finite-volume effects are well under control within current statistical uncertainties.
We also examined the $R$ dependence for other sets of $\{\Delta t_i,\Delta t_f,t_s\}$ and the same conclusion holds.

        \begin{figure}[htbp]
                \begin{center}
                        \includegraphics[width=85mm]{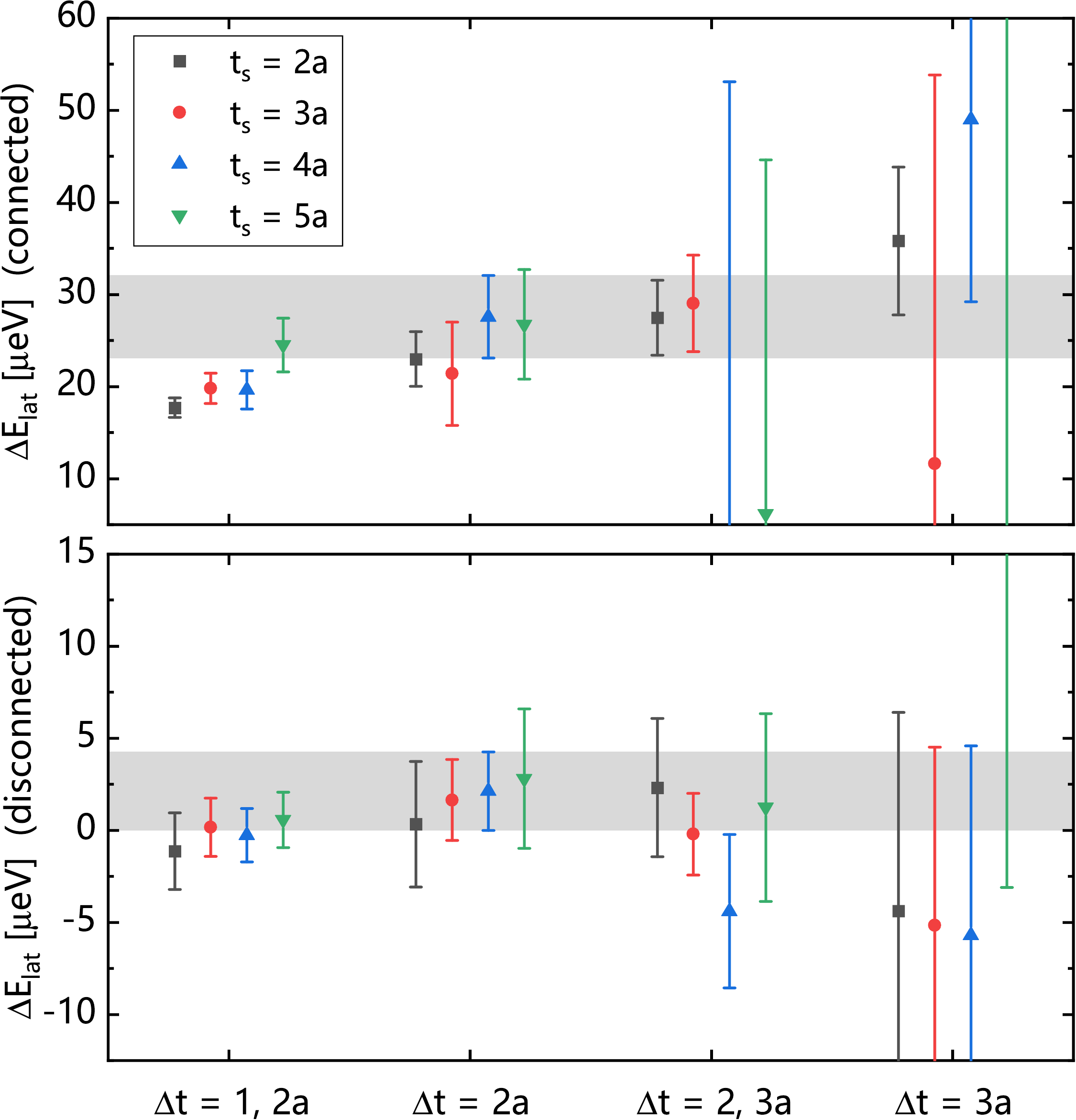}
                        \caption{Results of $\Delta E_{\lat}$ for multiple $\{\Delta t_i,\Delta t_f, t_s\}$. The results at different $t_s$ have been shifted horizontally for an easy comparison.}
                        \label{fig:TPE_ts}
                \end{center}
        \end{figure}
	
	The results of $\Delta E_{\lat}$ for different $\{\Delta t_i,\Delta t_f,t_s\}$
are shown in Fig.~\ref{fig:TPE_ts}.
For the connected part, we find that the result at $\{\Delta t_i,\Delta t_f,t_s\}=\{2a,2a,4a\}$ is well consistent with the ones
at $\{2a,2a,3a\}$ and $\{2a,2a,5a\}$. In addition, this result also agrees well with all the data at $\Delta t_{i/f}\ge 2a$.
For the disconnected part, the results for various $\{\Delta t_i,\Delta t_f,t_s\}$ are all consistent. The agreement with 0 suggests that
the disconnected contributions are relatively small.
We thus quote $\Delta E_{\lat}$ at $\{\Delta t_i,\Delta t_f,t_s\}=\{2a,2a,4a\}$ as the final result, and obtain
\be
\Delta E_{\lat}=
\begin{cases}
27.6(4.5)~\text{$\mu$eV}, & \text{connected part}, \\
2.1(2.1)~\text{$\mu$eV}, & \text{disconnected part}, \\
29.7(4.9)~\text{$\mu$eV}, & \text{total contribution}.
\end{cases}
\ee
The TPE correction is given by
        \be \label{eq:result}
        \Delta E_{\text{TPE}} = -28.9(4.9)~\mu\text{eV} + 93.72~\mu\text{eV}/\text{fm}^2\cdot \langle r_p^2 \rangle.
        \ee

Combining Eq.~(\ref{eq:result}) with (\ref{eq:muH_LS_theory}) and comparing the theoretical value with the experimental one, we obtain $\sqrt{\langle r_p^2\rangle}=0.84136(65)$ fm, which is consistent with $\sqrt{\langle r_p^2\rangle}=0.84087(39)$ fm quoted
from $\mu$H experiment. On the other hand, if we put the $\mu$H value of $\sqrt{\langle r_p^2\rangle}$ in Eq.~(\ref{eq:result}), we obtain $\Delta E_{\text{TPE}}=37.4(4.9)$ $\mu$eV, which agrees with the previous theoretical results ranging from 20 to 50 $\mu$eV.

We remark here that this calculation is performed at the nearly physical pion mass but with a relatively coarse lattice spacing $a=0.1943(8)$ fm.
We have used multiple $\{\Delta t_i,\Delta t_f, t_s\}$ to control the excited-state effects and examined the finite-volume effects by
studying the $R$ dependence. Thus 
we expect that the dominant systematic uncertainty arises from the lattice discretization effects. It is our future task to further control these effects using the ensembles with finer lattice spacings.

	\section{Conclusion} 
We have developed a method to calculate the TPE correction to the $\mu$H Lamb shift using lattice QCD. The methodology includes
\begin{itemize}
\item the derivation of the master formula~(\ref{eq:master_formula}) to remove IR divergence automatically and to compute the IR-finite $\Delta E$ using the hadronic functions $H_i(\vec{x},t)$ calculable from lattice QCD,
\item the design of an optimized subtraction scheme to significantly reduce finite-volume effects and statistical noise.
\end{itemize}
Using the new method, we perform a lattice calculation at $m_\pi = 142$ MeV. 
It demonstrates that lattice QCD can extend its horizon to study the important quantities relevant for atomic spectroscopy.
Within both statistical and systematic errors better controlled in the future, lattice studies can help answer more accurately 
the natural question - how large the proton is.

\begin{acknowledgements}

\section{Acknowledgments}

We gratefully acknowledge many helpful discussions with our colleagues from the
RBC-UKQCD Collaborations.
We thank M. Gorchtein and C.-Y. Seng for useful communications.
X.F., Y.F. and C.F.L. were supported in part by NSFC of China under Grants 
No. 12125501, No. 12070131001, and No. 12141501,
and National Key Research and Development Program of China under No. 2020YFA0406400. 
L.C.J. acknowledges support by DOE Office of Science Early Career Award No. DE-SC0021147
and DOE Award No. DE-SC0010339.
The calculation is carried out on Tianhe 3 prototype at Chinese National Supercomputer Center in Tianjin.
The computation is also performed under the ALCC Program of
the US DOE on the Blue Gene/Q (BG/Q) Mira computer at the Argonne Leadership Class Facility,
a DOE Office of Science Facility supported under Contract DE-AC02-06CH11357.
Computations for this work were carried out in part on facilities of the USQCD
Collaboration, which are funded by the Office of Science of the U.S. Department of Energy.

\end{acknowledgements}

	\bibliography{./ref.bib}

\clearpage

\setcounter{page}{1}
\renewcommand{\thepage}{Supplementary Information -- S\arabic{page}}
\setcounter{table}{0}
\renewcommand{\thetable}{S\,\Roman{table}}
\setcounter{equation}{0}
\renewcommand{\theequation}{S\,\arabic{equation}}
\setcounter{figure}{0}
\renewcommand{\thefigure}{S\,\arabic{figure}}

\section{Supplementary Material}

In this section, we expand on a selection of technical details.

\subsection{Construction of four-point correlation functions}

\begin{figure}[htbp]
                \begin{minipage}[h]{0.24\linewidth}
                        \centerline{ \includegraphics[width=50pt]{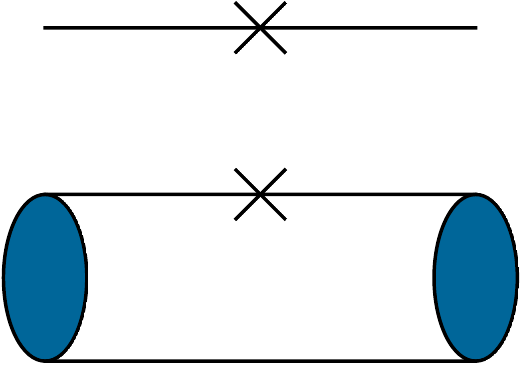} }
                        type 1
                \end{minipage}
                \begin{minipage}[h]{0.24\linewidth}
                        \centerline{ \includegraphics[width=50pt]{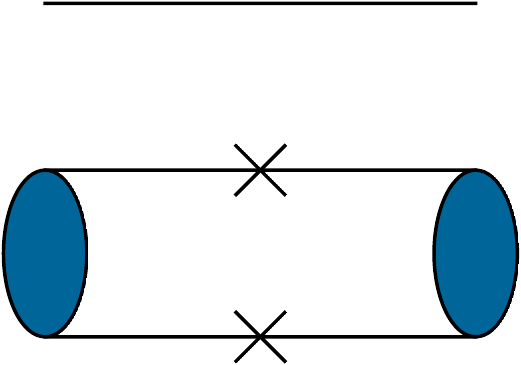} }
                        type 2
                \end{minipage}
                \begin{minipage}[h]{0.24\linewidth}
                        \centerline{ \includegraphics[width=50pt]{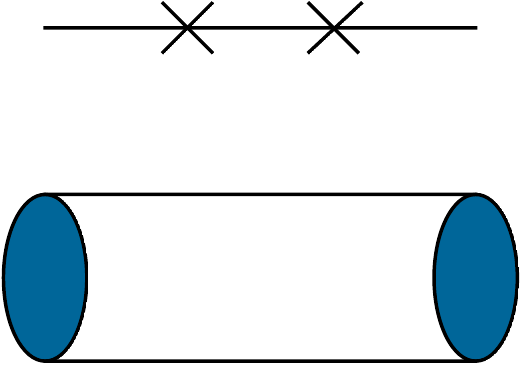} }
                        type 3
                \end{minipage}
                \begin{minipage}[h]{0.24\linewidth}
                        \centerline{ \includegraphics[width=50pt]{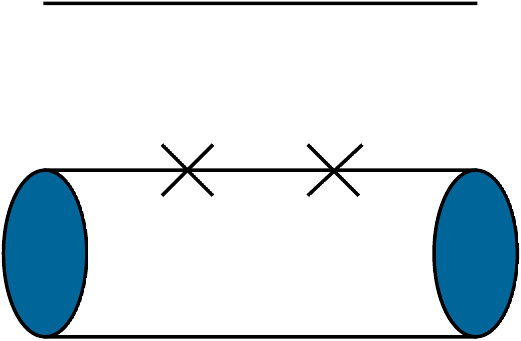} }
                        type 4
                \end{minipage}
                \begin{minipage}[h]{0.24\linewidth}
                        \centerline{ \includegraphics[width=50pt]{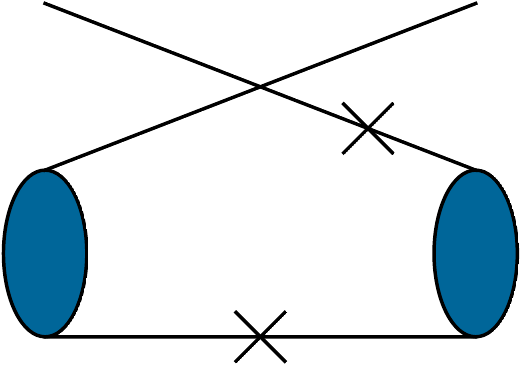} }
                        type 5
                \end{minipage}
                \begin{minipage}[h]{0.24\linewidth}
                        \centerline{ \includegraphics[width=50pt]{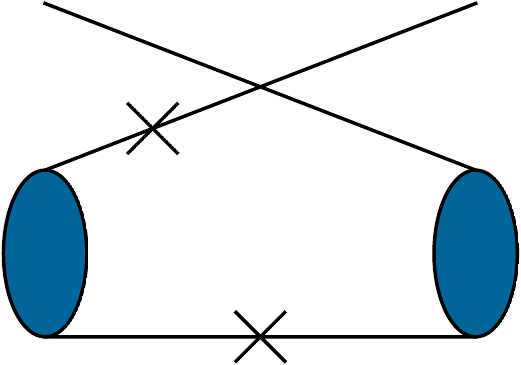} }
                        type 6
                \end{minipage}
                \begin{minipage}[h]{0.24\linewidth}
                        \centerline{ \includegraphics[width=50pt]{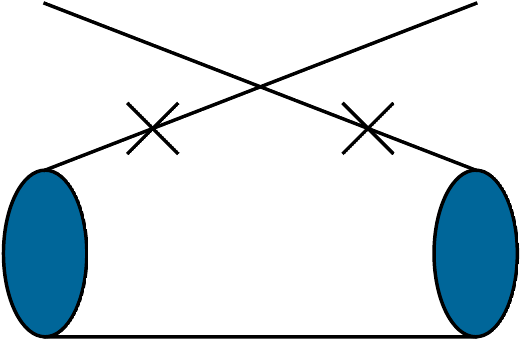} }
                        type 7
                \end{minipage}
                \begin{minipage}[h]{0.24\linewidth}
                        \centerline{ \includegraphics[width=50pt]{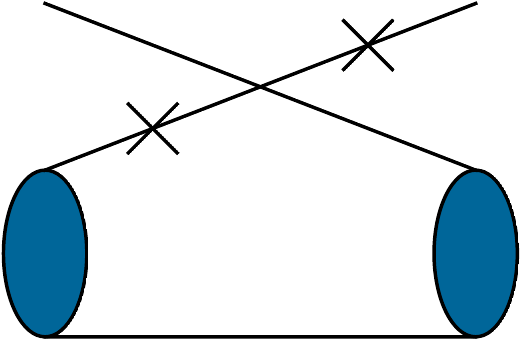} }
                        type 8
                \end{minipage}
                \begin{minipage}[h]{0.24\linewidth}
                        \centerline{ \includegraphics[width=50pt]{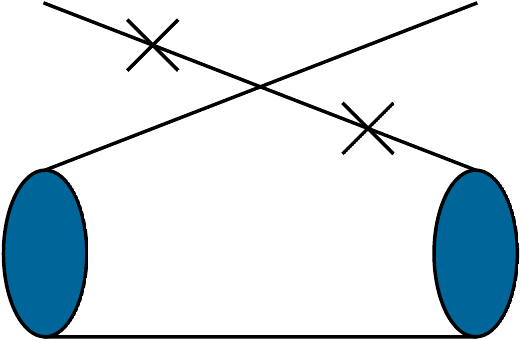} }
                        type 9
                \end{minipage}
                \begin{minipage}[h]{0.24\linewidth}
                        \centerline{ \includegraphics[width=50pt]{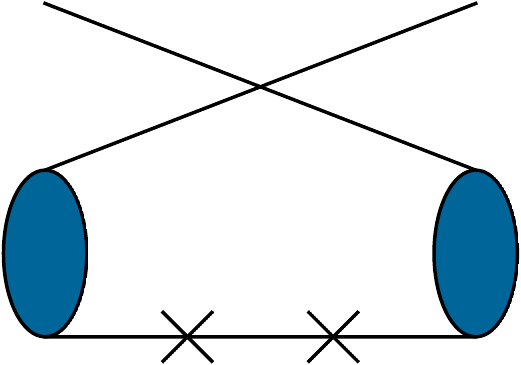} }
                        type 10
                \end{minipage}
                \caption{The connected diagrams for the four-point correlation function are classified into 10 types.
		The blob denotes the isospin $I=0$ and spin $S=0$ diquark.}
                \label{fig:connected_diag}
        \end{figure}

The quark contractions for the disconnected diagrams are relatively simple, while
the connected diagrams involve 10 different types of contractions as shown in Fig.~\ref{fig:connected_diag}.
The connected part of the hadronic function is composed of 10 contributions through
\be
H_{\mu\nu}^{\text{conn}}(\vec{x},t)=\sum_{n=1}^{10}c_n\,\left[H_{\mu\nu}^{(n)}(\vec{x},t)+H_{\nu\mu}^{(n)}(-\vec{x},-t)\right],
\ee
where the superscript $(n)$ indicates the contraction type and the coefficients $c_n$ are given by
\be
c_n=\left\{\frac{2}{9},-\frac{2}{9},\frac{4}{9},\frac{5}{9},\frac{2}{9},\frac{2}{9},-\frac{4}{9},-\frac{4}{9},-\frac{4}{9},-\frac{1}{9}\right\}.
\ee

        \begin{figure}[htbp]
                \begin{minipage}[h]{0.45\linewidth}
                        \centerline{ \includegraphics[width=50pt]{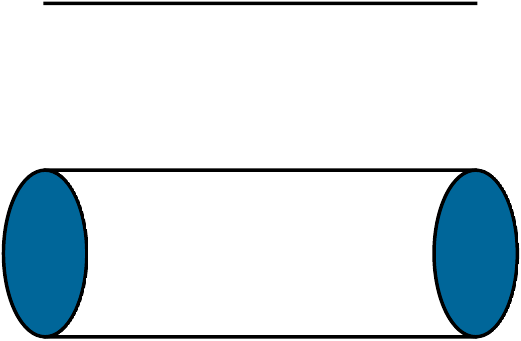} }
                        type 1
                \end{minipage}
                \begin{minipage}[h]{0.45\linewidth}
                        \centerline{ \includegraphics[width=50pt]{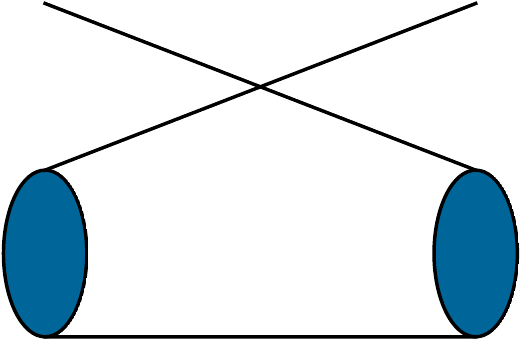} }
                        type 2
                \end{minipage}
                \caption{The two-point correlation function is constructed using two types of diagrams.}
                \label{fig:two_point_function}
        \end{figure}

For both connected and disconnected diagrams, the hadronic function can be extracted from the ratio of the correlation functions through
\be
\label{eq:ratio}
H_{\mu\nu}(x-y)=2M\frac{C_{\mu\nu}(x,y;t_i,t_f)}{C_2(t_i,t_f)},
\ee
where
\ba
&&C_{\mu\nu}(x,y;t_i,t_f)
\nn\\
&=&\sum_{\vec{x}_f, \vec{x}_i}\operatorname{Tr}[ \mathcal{P} \langle \psi_p(\vec{x}_f,t_f) j_\mu(x) j_\nu(y) \psi_p^\dagger(\vec{x}_i,t_i)\rangle],
\ea
and
\be
C_2(t_i,t_f)=\frac{1}{L^3}\sum_{\vec{x}_f, \vec{x}_i}\operatorname{Tr}[ \mathcal{P} \langle \psi_p(\vec{x}_f,t_f) \psi_p^\dagger(\vec{x}_i,t_i)\rangle].
\ee
Here, the two-point correlation function $C_2(t_i,t_f)$ is constructed using the two types of diagrams shown in Fig.~\ref{fig:two_point_function}
\be
C_2(t_i,t_f)=C_2^{(1)}(t_i,t_f)-C_2^{(2)}(t_i,t_f).
\ee

In practice, for each configuration we calculate point-source propagators at $N=1024$ random spacetime locations. We place the nucleon creation and annihilation operators as well as
one vector current at these random locations and treat the location of the other vector current as the sink. In this way, we are able to compute the hadronic
function $H_{\mu\nu}(\vec{x},t)$ at arbitrary $(\vec{x},t)$ and multiple sets of the time slices $\{t_i,t_f\}$ (or equivalently $\{\Delta t_i,\Delta t_f\}$). The spatial volume summation over $\vec{x}_i$ and $\vec{x}_f$ is replaced by the random points summation multiplied with a normalization factor.

\subsection{Numerical confirmation of Eq.~(\ref{eq:low_mom_relation})}

        \begin{figure}[htbp]
                \begin{center}
                        \includegraphics[width=80mm]{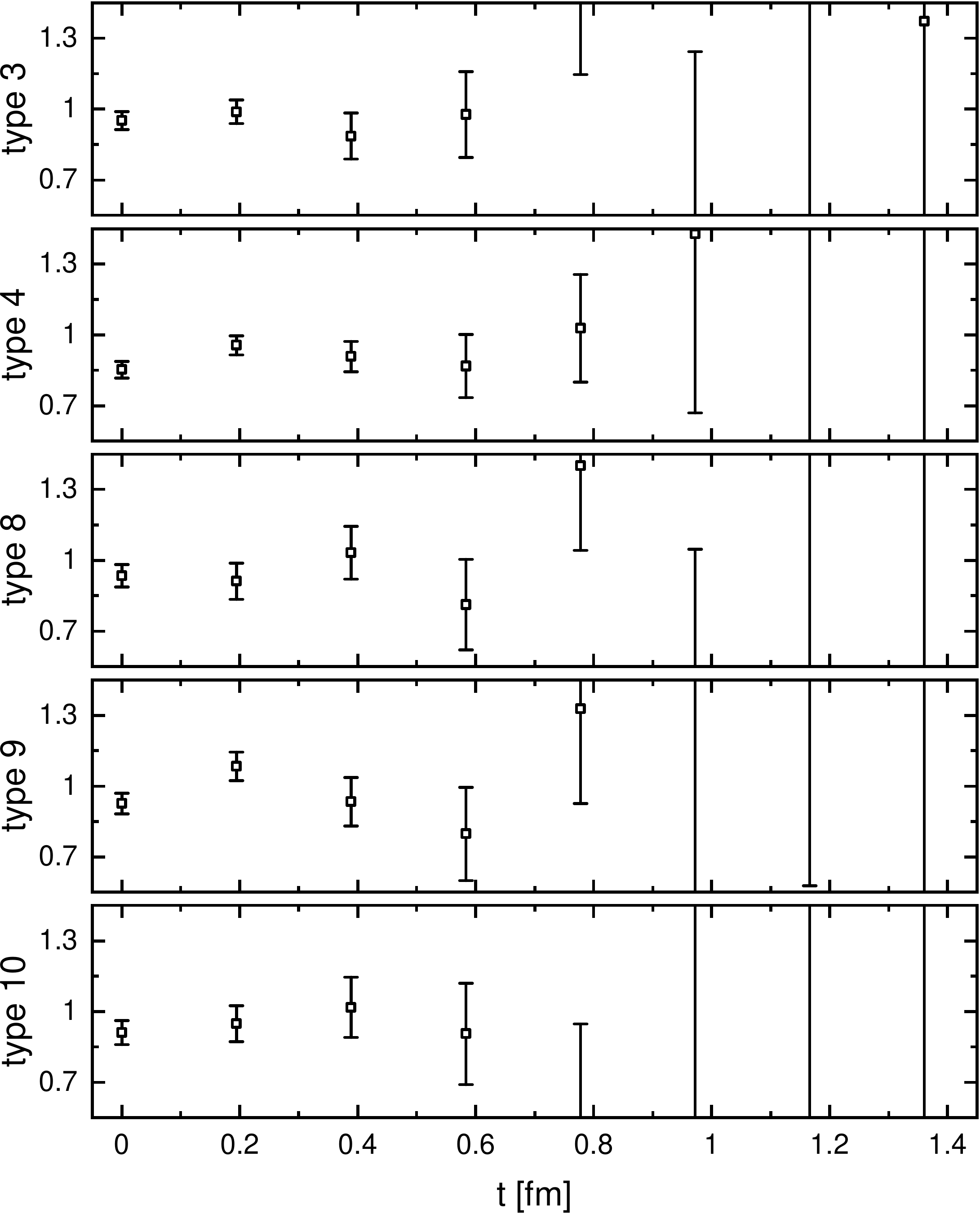}
                        \caption{Numerical confirmation of the charge conservation, i.e. Eq.~(\ref{eq:one_line}), using the lattice data at $\{\Delta t_i,\Delta t_f\}=\{2a,2a\}$.}
                        \label{fig:charge_conservation_one_line}
                \end{center}
        \end{figure}

       \begin{figure}[htbp]
                \begin{center}
                \includegraphics[width=80mm]{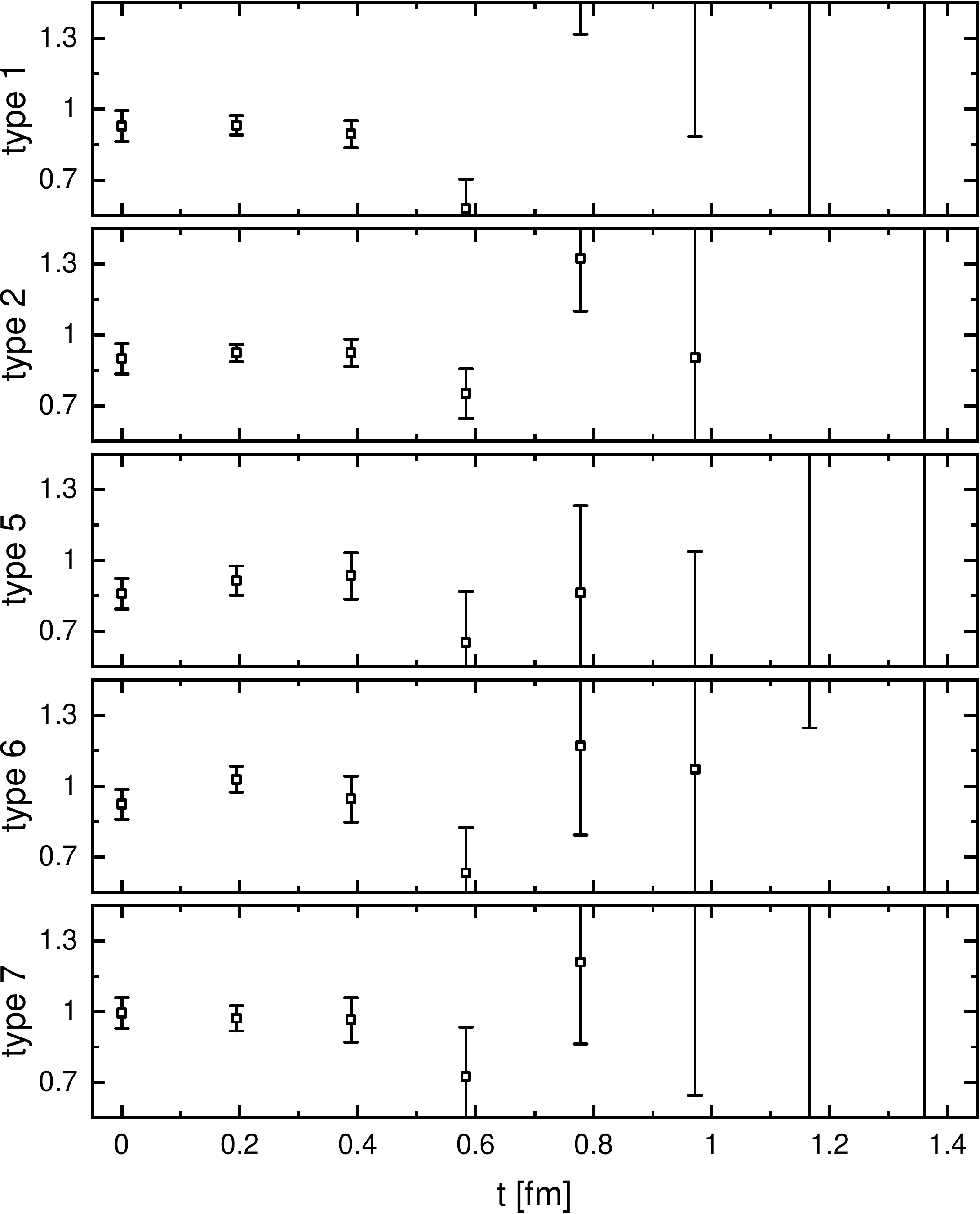}
                        \caption{Numerical confirmation of the charge conservation, i.e. Eq.~(\ref{eq:two_lines}), using the lattice data at $\{\Delta t_i,\Delta t_f\}=\{2a,2a\}$.}
                        \label{fig:charge_conservation_two_lines}
                \end{center}
        \end{figure}

        \begin{figure}[htbp]
                \begin{center}
                        \includegraphics[width=80mm]{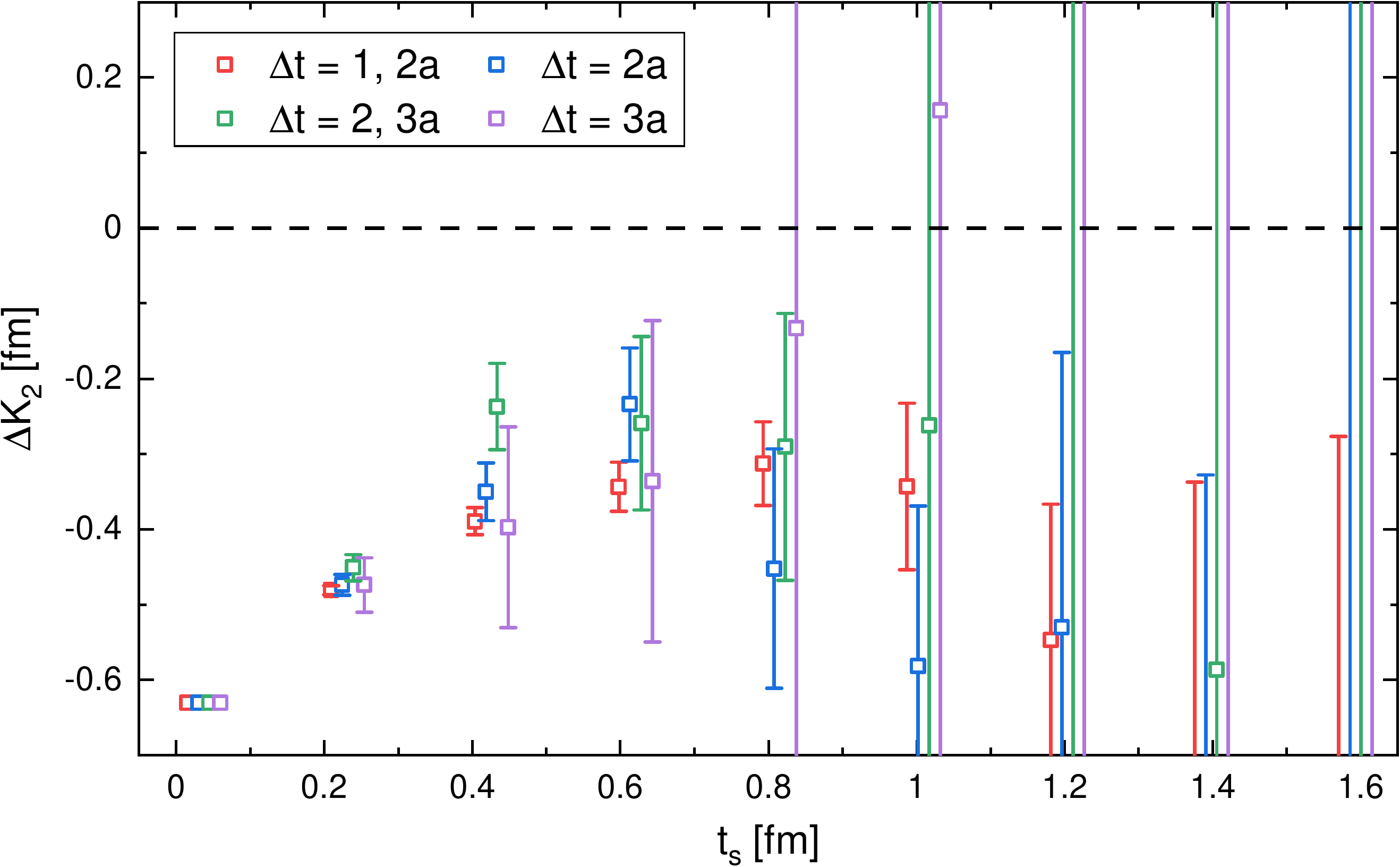}
                        \caption{Examination of $\Delta K_2\equiv K_2-\frac{3}{M}=0$ at sufficiently large $t_s$. The connected part of the data is used here.}
                        \label{fig:K2}
                \end{center}
        \end{figure}

Eq.~(\ref{eq:low_mom_relation}) includes two results: $K_1=2\,t_s$ and $K_2=\frac{3}{M}$. The former is equivalent to
the condition of charge conservation, which requires
\be
\label{eq:charge_conservation}
\frac{1}{2M}\int \ud^3\vec{x}\,\langle p|\operatorname{T}[j_0(\vec{x},t)j_0(0)]|p\rangle=1
\ee 
at arbitrary $t$.
As the charge conservation holds for all the connected diagrams, in Figs.~\ref{fig:charge_conservation_one_line} and \ref{fig:charge_conservation_two_lines} we show the examination of Eq.~(\ref{eq:charge_conservation})
for each individual type of diagram. For these diagrams with two currents inserted into a single quark line, namely type 3, 4, 8, 9 and 10, we examine the ratio of
\be
\begin{cases}
\label{eq:one_line}
\sum_\vec{x}\frac{C_{00}^{(n)}(x,y;t_i,t_f)+C_{00}^{(n)}(y,x;t_i,t_f)}{C_2^{(1)}(t_i,t_f)}=1, & n=3,4 \\
\sum_\vec{x}\frac{C_{00}^{(n)}(x,y;t_i,t_f)+C_{00}^{(n)}(y,x;t_i,t_f)}{C_2^{(2)}(t_i,t_f)}=1, & n=8,9,10.
\end{cases}
\ee
For these diagrams with two currents inserted into two different quark lines, namely type 1, 2, 5, 6 and 7, we examine the ratio of 
\be
\begin{cases}
\label{eq:two_lines}
\sum_\vec{x}\frac{C_{00}^{(n)}(x,y;t_i,t_f)+C_{00}^{(n)}(y,x;t_i,t_f)}{2C_2^{(1)}(t_i,t_f)}=1, & n=1,2 \\
\sum_\vec{x}\frac{C_{00}^{(n)}(x,y;t_i,t_f)+C_{00}^{(n)}(y,x;t_i,t_f)}{2C_2^{(2)}(t_i,t_f)}=1, & n=5,6,7.
\end{cases}
\ee
Note that in Eq.~(\ref{eq:two_lines}) there is an additional factor of $2$ in the denominator.
As far as the charge conservation is confirmed, for the connected part we can construct a ratio of
\be
H_{\mu\nu}^{\text{conn}}(x-y)=2M\frac{C_{\mu\nu}^{\text{conn}}(x,y;t_i,t_f)}{\sum_\vec{x}C_{00}^{\text{conn}}(x,y;t_i,t_f)}.
\ee
It helps reduce the statistical uncertainty by a factor of $1.5$ when compared to the usage of Eq.~(\ref{eq:ratio}).

To examine the second condition of $K_2=\frac{3}{M}$, in Fig.~\ref{fig:K2} we make a plot of $\Delta K_2\equiv K_2-\frac{3}{M}$ as a function of $t_s$ for multiple 
$\{\Delta t_i,\Delta t_f\}$. When $t_s$ becomes sufficiently large, we expect to have $\Delta K_2=0$. However,
the result at $\{\Delta t_i,\Delta t_f,t_s\}=\{2a,2a,4a\}$ is not fully consistent with $0$, suggesting that the residual
systematic effects such as the lattice artifacts need to be further controlled in the future work.

\subsection{Efficiency of the optimized subtraction scheme}

\begin{figure}[htbp]
                \begin{center}
                        \includegraphics[width=80mm]{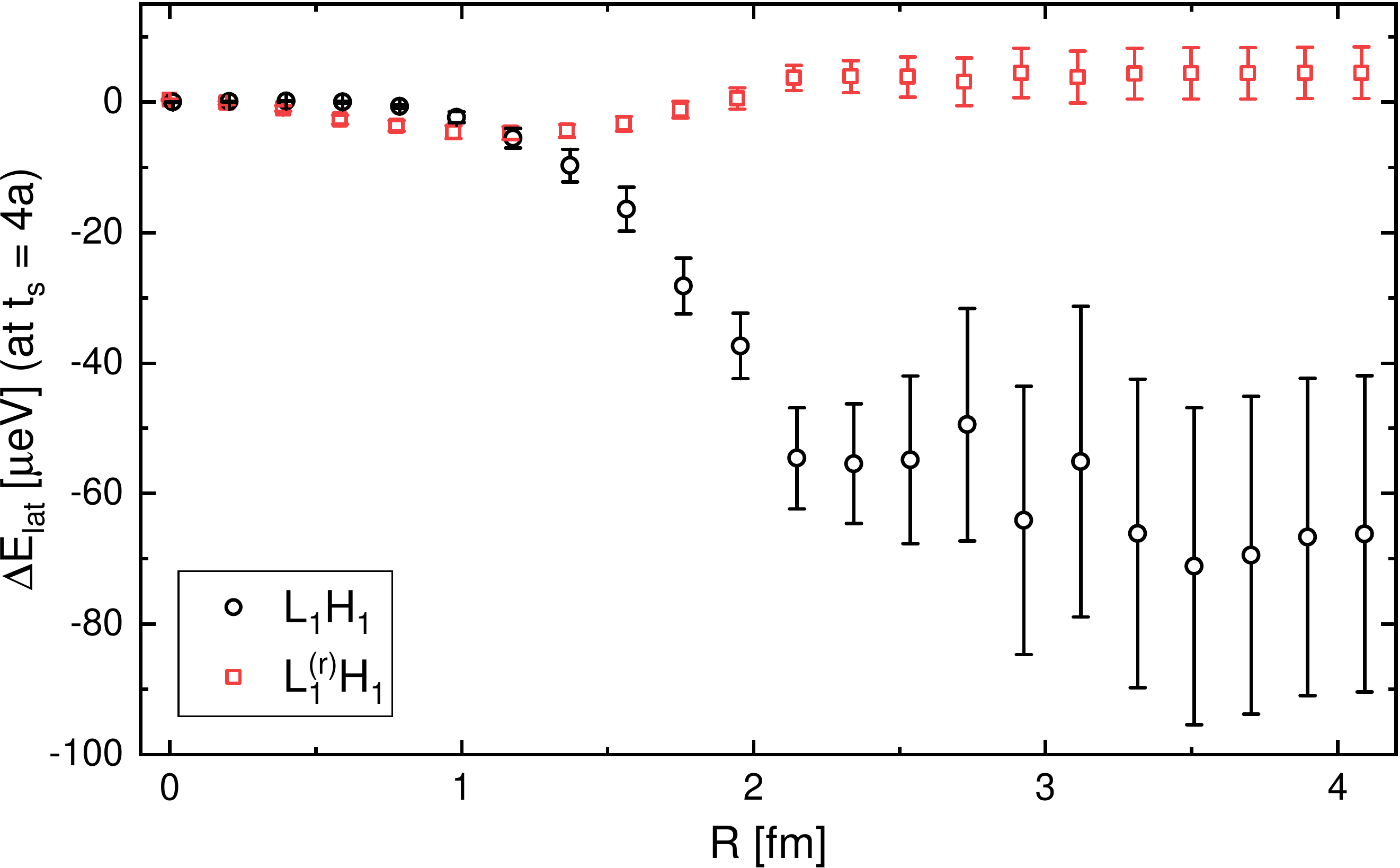}
                        \caption{Comparison of the lattice results using or not using the optimized subtraction scheme. We use the connected part of the results computed at
$\{\Delta t_i,\Delta t_f,t_s\}=\{2a,2a,4a\}$.}
                        \label{fig:L1_or_L1r}
                \end{center}
        \end{figure}

By using the optimized subtraction scheme, the lattice results are significantly improved. In Fig.~\ref{fig:L1_or_L1r}, we compare the lattice results using or not using the 
optimized subtraction scheme and find that the former gain a precision $\sim$6 times better.

\end{document}